\documentclass[fleqn,usenatbib]{mnras}
\usepackage{newtxtext,newtxmath}
\usepackage[T1]{fontenc}

\DeclareRobustCommand{\VAN}[3]{#2}
\let\VANthebibliography\thebibliography
\def\thebibliography{\DeclareRobustCommand{\VAN}[3]{##3}\VANthebibliography}
\DeclareRobustCommand{\DE}[3]{#2}
\let\DEthebibliography\thebibliography
\def\thebibliography{\DeclareRobustCommand{\DE}[3]{##3}\DEthebibliography}

\usepackage{graphicx}	% Including figure files
\usepackage{amsmath}	% Advanced maths commands

\def\toprule{\hline}
\def\botrule{\\\hline}

\title[sdBs with COMPAS: the Galactic Population]{Population synthesis of hot subdwarf B stars with COMPAS: on the observed Galactic population}

\author[N. Rodriguez-Segovia \& A. J. Ruiter]{
Nicol\'{a}s Rodr\'{i}guez-Segovia,$^{1}$\thanks{E-mail: nj.rsegovia@gmail.com (NRS)}
and Ashley J. Ruiter$^{1,2}$
\\
$^{1}$School of Science, University of New South Wales, Australian Defence Force Academy, Canberra, ACT 2600, Australia\\
$^{2}$OzGrav: The ARC Centre of Excellence for Gravitational Wave Discovery, Hawthorn, VIC 3122, Australia
}

\date{Accepted XXX. Received YYY; in original form ZZZ}

\pubyear{\the\year{}}

\begin{document}
\label{firstpage}
\pagerange{\pageref{firstpage}--\pageref{lastpage}}
\maketitle

\begin{abstract}
Hot subdwarf B stars (sdBs) are helium-burning stars with thin hydrogen-rich envelopes. Their most widely accepted formation channels involve binary evolution and progenitors near the tip of the red giant branch, thus studying these objects improves our knowledge of complicated astrophysical processes such as common envelope evolution and the helium flash. In this work, we compare the observed sdB population against a synthetic Galactic population generated through the binary population synthesis code \textsc{compas}, which allows us to estimate the physical properties of the current-day Galactic sdB population. We show that our synthetic sdB population matches the general properties of the observations quite well in the Kiel diagram when either a normal or log-normal distribution is assumed for the assignment of hydrogen-rich envelope masses. We also find that the canonical mass assumption should only be confidently assumed for specific system configurations and that the estimated number of sdBs found within 500 pc of the Sun in our model is at least four times higher than the observational one. We recover the observational P-q relation for sdBs plus main sequence companions, while a similar relation between sdBs and helium white dwarf companions is rather complicated. We conclude that a better understanding of hydrogen-rich envelopes is needed, as well as an observational characterization of the sdB plus main sequence companions earlier than spectral type $\sim$\,F. These issues aside, atmospheric properties, companion types, period and mass distributions are in good agreement with observational and theoretical studies available in the literature.
\end{abstract}


%%%%%%%%%%%%%%%%%%%%%%%%%%%%%%%%%%%%%%%%%%%%%%%%%%

%%%%%%%%%%%%%%%%% BODY OF PAPER %%%%%%%%%%%%%%%%%%

\section{Introduction}

Subdwarf B (sdB) stars are understood as core helium-burning objects that have lost most of their hydrogen-rich mass and are left with a low-mass outer layer of up to 0.02$M_\odot$ \citep{Heber1986}. They are usually described as having a \textit{canonical mass} of $\sim 0.47 M_\odot$ due to their origin as helium cores of stars nearing helium ignition \citep[][and references therein]{Heber2016, Heber2024}. This assumption originates from the preference of the IMF for low-mass stars \citep[e.g.,][section 2.2 and figure 1]{Hopkins2018}, coupled with stellar evolution models showing a mass plateau for the helium core around the tip of the red giant branch (RGB) for these stars \citep[e.g.,][figure C3]{Ostrowski2021}. However, it must be clarified that theory also allows for a wider variety of masses, either due to progenitors whose cores are not completely degenerate or the mass loss event that gives birth to the sdB being triggered further away from helium ignition \citep[e.g.,][]{Han2002,ArancibiaRojas2024,Rodriguez2024}.

The observed and predicted characteristics of sdB stars make them interesting objects in a wide variety of fields. Namely, they have been observed to pulsate, which makes them targets for asteroseismologic studies (e.g., Murphy \citeyear{Murphy2018} for the utility of pulsating sdBs in binary systems, or Charpinet et al. \citeyear{Charpinet2013} for an asteroseismology-focused review), while their high binary fraction \citep{Pelisoli2020} and predicted formation channels \citep[e.g.,][]{Mengel1976,Han2002, Rodriguez2024} make them ideal laboratories for the study of binary evolution and related processes \citep[such as common envelope events, e.g.,][]{DeMarco2011}. Further, they have been used in studies of the UV excess in elliptical galaxies \citep{Podsiadlowski2008} and have recently been highlighted as potential tracers of Galactic evolutionary history \citep{Vos2020}. There have also been studies focused on groups of sdBs with particular characteristics, such as the case of gravitational wave sources (including a LISA verification source, see Kupfer et al. \citeyear{Kupfer2024}), chemically peculiar sdBs \citep{Naslim2011, Naslim2013} and high-velocity stars \citep[][and references therein]{Geier2024}.

To study the formation channels of sdB stars and revisit the widely known results of \citet{Han2002}, \citet{Rodriguez2024} (Paper I hereafter) performed a parameter study by using the rapid binary population synthesis (BPS) code Compact Object Mergers: Population Astrophysics and Statistics \citep[\textsc{compas}, Team COMPAS:][]{Riley2022}. Key components of Paper I include the addition of hydrogen-rich envelope remnants to the helium main sequence (HeMS) stars within the \citet{Hurley2000} evolutionary scheme inherited by \textsc{compas}, based on fits to the detailed stellar models from \citet{Bauer2021}. This allows for improved agreement between observables and population synthesis predictions of atmospheric parameters (surface gravity and effective temperature). In this work we expand on what was presented in Paper I by implementing a Galaxy-like model in terms of metallicity and total stellar mass, which allows a direct comparison between the observed current-day Galactic sdB population and the results from our simulations.

This work is structured as follows: Section \ref{sect:met} contains the set of methods used to perform this study (construction of the initial synthetic populations, re-sampling, and a theoretical analysis of the orbital period -- mass ratio relation), section \ref{sect:res} shows our results (BPS predictions on the current number of sdBs, relevance of different channels, atmospheric properties, comments on the orbital period -- mass ratio relation, mass and period distributions), and section \ref{sect:conc} presents our summary and conclusions.

\section{Methods}\label{sect:met}

\subsection{Synthetic Population}\label{sect:synth}

We follow the same approach as in Paper I, using the fits made to the \citet{Bauer2021} \textsc{mesa} models that allow us to transform naked HeMS stars crossing the pre-defined \textit{sdB box} (equations 1--4 in Paper I) into sdB candidates. We continue using \textsc{compas} as our base BPS implementation, though we use \texttt{v03.09.00} which includes updated critical mass ratios from \citet{Ge2024b}, among other updates (Team COMPAS: Riley et al., in prep.). The same methods to incorporate candidates whose mass values are below the expected core mass at helium ignition have been employed, and candidates born from mergers are dealt with under the same assumptions. For these details, the reader is referred to Paper I. In this new work, however, we intend to mimic the Galactic sdB population and, therefore, we must consider a series of changes. These include fixing the values used for common envelope efficiency \citep[$\alpha = 0.2$,][]{Zorotovic2010}, no accretion by the companion during mass transfer episodes (usually labelled as $\beta = 0$), and mass lost from the system under the \textit{isotropic re-emission} scheme from the vicinity of the accretor\footnote{For consistency with Paper I, this implies setting the \textit{MacLeod linear fraction} coefficients to 0.} \citep{Vos2020}. Additional changes related to the sampling methodology are detailed below.

\begin{table}
\centering
\caption{Details of the implementation of the Besan\c{c}on model of the Galaxy \citep{Robin2003} used in this work. Updated density weights ($\rho_0$) are taken from \citet{Czekaj2014}, while mass fractions ($f_i$) are computed by integrating the radial profiles of each component.}\label{tab:besancon}
{\begin{tabular}{@{\extracolsep{\fill}}lcccc}
\toprule
Component & Age & $\left[\frac{\rm Fe}{\rm H}\right]$ & $\rho_0$ & $f_i$\\
  & (Gyr) & (dex) & ($10^{-3}M_\odot\, {\rm pc}^{-3}$) & (\%) \\
\hline
Thin Disk 1 & 0-0.15 & 0.01 $\pm$ 0.12 & 1.89 &  0.88  \\
Thin Disk 2 & 0.15-1 & 0.03 $\pm$ 0.12 & 5.04 &  4.63  \\
Thin Disk 3 & 1-2 & 0.03 $\pm$ 0.10 & 4.11 &  5.23  \\
Thin Disk 4 & 2-3 & 0.01 $\pm$ 0.11 & 2.84 &  5.28  \\
Thin Disk 5 & 3-5 & -0.07 $\pm$ 0.18 & 4.88 &  11.44  \\
Thin Disk 6 & 5-7 & -0.14 $\pm$ 0.17 & 5.02 &  13.26  \\
Thin Disk 7 & 7-10 & -0.37 $\pm$ 0.20 & 9.32 &  24.81  \\
Bulge  & 8-10 & 0.00 $\pm$ 0.40 &  &  9.89  \\
Thick Disk & 10 & -0.78 $\pm$ 0.30 & 2.91 &  22.4  \\
Halo & 14 & -1.78 $\pm$ 0.50 & 9.2$\times 10^{-3}$ &  2.18  
\botrule
\end{tabular}}
\end{table}

\begin{figure*}
\includegraphics[width=1\linewidth]{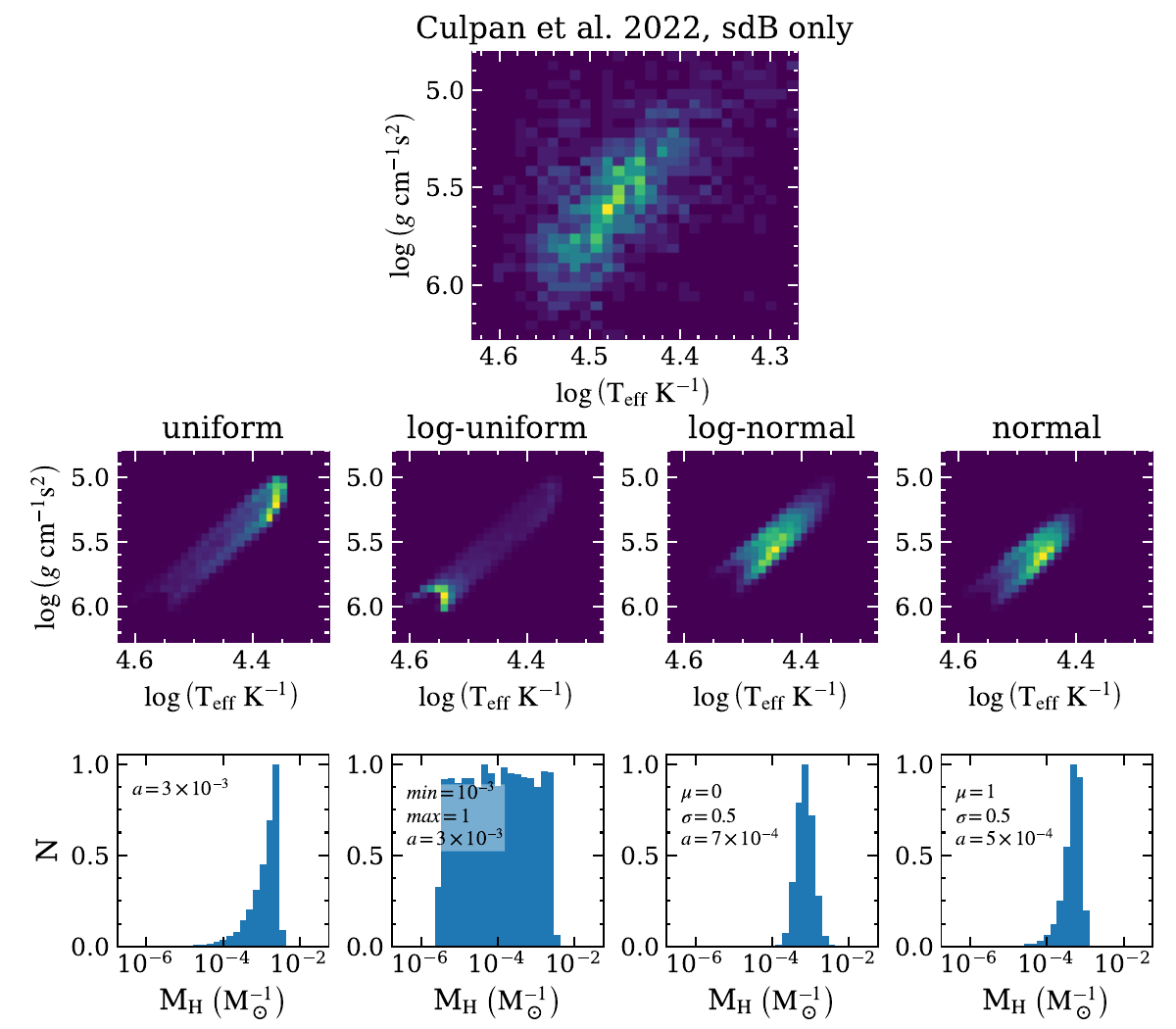}
\caption{Graphic comparison between the \citet{Culpan2022} observed sdB population (top) and a fiducial population composed of 10,000 sdBs with masses randomly sampled from the 0.47 -- 0.49 $M_\odot$ range (middle row). Different assumed distributions for the hydrogen-rich mass envelope are represented in each column of the middle row, with the actual normalized distributions being shown in the bottom row. From left to right: \texttt{numpy.random.default\_rng().random}, \texttt{scipy.stats.loguniform.rvs}, \texttt{numpy.random.default\_rng().lognormal} and \texttt{scipy.stats.truncnorm.rvs} \citep{Harris2020, Virtanen2020}. Required parameters are shown in each panel, except for $a$ which corresponds to a multiplicative factor applied to resulting arrays.}
\label{fig:env_test}
\end{figure*}

An adequate Galactic model must take into account the different structural components, stellar populations, and evolutionary history of the Galaxy, which results in a fixed metallicity value no longer being an appropriate choice. For these reasons, we have opted to use the Besan\c{c}on Galactic model \citep{Robin2003}, which provides temporal, spatial, kinematic and chemical (iron abundance) distributions for a set of predefined components that mimic the observed Galaxy. We summarize the properties of this model in Table \ref{tab:besancon}. \citet{Robin2003} explain that the iron abundances are modelled as normal distributions with a mean equal to the provided value per Galactic component, and standard deviation following the reported uncertainties. We can randomly draw iron abundances from these distributions, which are then transformed to metallicity (Z) values by assuming a relation with the nominal Solar Z value (equation 7 in Paper I), though some of the resulting metallicities are beyond the interpolation limits in \textsc{compas}. In these cases, we map the metallicities to the closest boundary, either the minimum (10$^{-4}$) or maximum (0.03) possible Z.

In Paper I we sampled around 280,000 systems from the \citet{Moe2017} orbital parameter distributions (as implemented in the \texttt{sampleMoeDiStefano.py} script included with \textsc{compas}), with masses in the 0.08 -- 150 $M_\odot$ range. Here, however, our initial population consists of 1,600,000 binaries sampled from the same distributions. This increase adds variety to our final population, as it increases the number of unique binary system configurations available to draw from during the re-sampling steps described in section \ref{sect:gal}. It should also be noted that since the sampler aims to reproduce the statistics from \citet{Moe2017}, it readily incorporates the inner binaries of triple systems. This means that the final numbers from our model would be similar to what can be found from a "hierarchical triple" population synthesis, while the relative relevance of each formation channel could vary \citep[according to the results from][]{Preece2022}.

A final remark is that in this work we have also changed how hydrogen-rich envelopes are sampled. While we still randomly generate values in the range 0--3$\times 10^{-3}\,\,M_\odot$, these are no longer drawn from a uniform distribution. Instead, values are drawn from a log-normal distribution with mean equal to 0 and standard deviation equal to 0.5. The resulting array is then multiplied by $7\times10^{-4}$ to make it coincide with the available range of hydrogen-rich envelope masses from the \citet{Bauer2021} models, while any values outside this range are forcefully mapped to the closest bound in a similar way to what was done with metallicities. This configuration was selected after the observational sdB sample from \citet{Culpan2022} was compared against several different hydrogen-rich envelope mass distributions tested in a fiducial synthetic population of 10,000 sdB stars with masses in the range 0.47 -- 0.49$\,\,M_\odot$ (around canonical mass) born from low-mass progenitors. These stars were also assigned random evolutionary stages (fraction of the total time expected for core helium burning, the ${\rm t}_{\rm r}$ variable in Paper I) drawn from a uniform distribution. While our approach does not provide a direct comparison between observed and predicted properties of the Galactic sdB population, it does yield an initial estimate of the distribution of sdBs in the Kiel diagram, under the assumption that most observed sdBs are indeed born from low-mass progenitors (and therefore have a mass close to the canonical value). Figure \ref{fig:env_test} depicts the resulting comparison between the tested envelope mass distributions and the observational sample from \citet{Culpan2022}. The figure implies that hydrogen-rich envelope masses uniformly distributed (either in linear or logarithmic space) are not a good proxy of the current observed sdB population, since they result in a sample that clusters at extreme surface gravities and temperatures, unlike the observed population that seems to be concentrated at intermediate values. Note, however, that this envelope distribution heavily relies on the assumption that the observed sdB distribution in the Kiel diagram is mostly due to the effect of changes in total mass and hydrogen-rich mass within the population, while we have not considered the influence of chemical composition due to limitations on the available \citet{Bauer2021} models. The chemical composition is one of the variables known to affect stellar properties and evolution due to its impact on the equation of state, nuclear energy generation and radiative opacity, to name a few \citep[e.g.,][]{Kippenhahn2012, Xin2022}; which would result in changes to effective temperature as well as surface gravity, effectively modifying the distribution of synthetic populations in the Kiel diagram.

\subsection{Galaxy-like Sampling}\label{sect:gal}

We have assumed a Galactic stellar-mass $M_{MW} = 6.43\times10^{10} \rm{M}_\odot$ \citep{McMillan2011}, and that the corresponding mass fraction contained in binary systems is 0.7, as computed by taking $M_{\rm binary} / (M_{\rm single}+M_{\rm binary})$ from our \textsc{compas} sample. We note that this is a simplified assumption since the binary fraction is known to depend on primary mass. The method described by \citet{Moe2017} shows an approach to such dependence (their equation 29 and figure 39), though the mass fraction contained in binary systems (different from the fraction of binary systems) also depends on the choice of IMF, from which primary masses are drawn. The final binary mass fraction can then be computed by choosing a prescription for mass ratios. However, since all these elements are accounted for in the \textsc{compas} sampler based on \citet{Moe2017}, we assume that the results for a relatively big population describe the sought binary mass fraction.

The result of the procedure described above is that each component of the Galaxy in our implementation of the Besan\c{c}on model contains the following mass:

\begin{equation}
    M_{i} = \frac{0.7M_{MW}f_i}{s},\label{eq:mass_component}
\end{equation}

\noindent where $M_i$ corresponds to one of the 10 components of the Galactic model, $f_i$ is the corresponding mass fraction of a given Galactic component, and $s$ corresponds to a linear factor that we set as 10. The introduction of this parameter is due to the linear scaling found in results from test runs when increasing the total mass of the Galaxy from $6.43\times10^{6} \rm{M}_\odot$ up to the previously defined $M_{MW}$ value by powers of 10. This choice saves computational resources and does not affect the final statistics. For $f_i$, one must integrate the radial profiles presented by \citet{Robin2003} while considering the local densities $\rho_0$ from \citet{Czekaj2014} as normalization constants. The results are presented in Table \ref{tab:besancon} and were computed using the Galactic population script from the LISA Synthetic UCB Gatalog project (Bobrick et al. 2025, in prep; Breivik et al. 2025, in prep)\footnote{The script is available online at \href{https://github.com/Synthetic-UCB-Catalogs/Galaxy-scripts}{https://github.com/Synthetic-UCB-Catalogs/Galaxy-scripts}}.

Taking the resulting $M_i$ values as the target mass for each component, we then proceed to randomly sample binary systems from the initial un-evolved population, until the total sampled mass is within 1\% of the corresponding $M_i$. Once this happens, only systems where at least one component goes through the HeMS phase during its lifetime are kept. These systems are also assigned a birth time that is either equal to the \textit{Age} column in Table~\ref{tab:besancon} (burst star formation) or randomly sampled from a uniform distribution between the given age range in the same column (constant star formation). The spatial distribution of stars in our synthetic Galaxy is then determined by the same density profiles that were used when computing the component mass fractions. 

A final step is to extract the \textit{current-day} sdB population, which we define as all the candidates that are sdB stars at a sample age of $13.95 \pm 0.05$ Gyr. This age was set in line with the oldest component (Halo) available in the Besan\c{c}on model. The 0.05 Gyr uncertainty range is motivated by the typical $\sim\!100$ Myr\footnote{However, sdBs with masses around 0.3$\,M_\odot$ can live longer, between 700 and 1200 Myr (depending on metallicity and hydrogen-rich envelope mass).} lifetime of HeMS stars \citep[e.g.,][]{Yungelson2008,ArancibiaRojas2024,Rodriguez2024}, which translates into considering only those HeMS stars that correspond to sdB-candidates in a system between 13.9 and 14 Gyr old, based on the sum of their assigned birth time during sampling, the time it takes to evolve into the HeMS stage, and the time that the adopted fits to the \citet{Bauer2021} models take to enter the \textit{sdB box}. An equivalent set of criteria must be considered for mergers, though these make use of the time it takes for a double helium white dwarf (WD) system to form instead of the time it takes a star to reach the HeMS stage, and also include the necessary time for the system to coalesce \citep[by using the work of][]{Peters1964} as done in Paper I.

\subsection{The Period -- Mass Ratio Relation}\label{sect:pq}

The orbital period -- mass ratio (P-q hereafter) and orbital period -- core mass relations for binary systems have been widely explored in the literature \citep[e.g.,][]{Rappaport1995,deKool1990,vanderSluys2006,Chen2013, Vos2020, Zhang2021,Ge2024}, and are particularly useful when exploring the characteristics of systems composed of an sdB and a low-mass MS star \citep[e.g.,][]{Chen2013, Vos2020} or an sdB and a helium white dwarf \citep[HeWD,][]{Zhang2021}. The reader is referred to such works for details about the derivation and applicability of these relations, and is warned that the relation can be also expressed in more specific forms, either the P -- \textit{companion mass} or P -- \textit{sdB mass} relation. In this work, we limit ourselves to providing some general context for the applicability of the P-q relation in the case of sdBs born through the stable mass transfer and common envelope scenarios. We highlight that its applicability in the context of BPS (section \ref{sect:pqres}) is straightforward, while it is not the case for the observational application. Still, the theoretical approach is useful when it comes to evaluating the methods and usage of such relations.

\subsubsection{Stable Mass Transfer}\label{sect:stable}

This particular case refers to systems where the sdB is born through a single stable Roche lobe overflow (RLOF) episode and typically results in long-period systems (from a few hundred up to $\sim 3$ thousand days) with main sequence (MS) companions (section 3.2.1 of Paper I, but also see Han et al. \citeyear{Han2002,Han2003} and more recently Vos et al. \citeyear{Vos2020}). The key assumption here is that the sdB progenitor follows the core mass--radius relation \citep[e.g.,][]{Rappaport1995} during the red giant branch (RGB). This assumption is acceptable, as long as the sdB progenitors from this channel possess a degenerate core before helium ignition\footnote{A relation does exist for progenitors that ignite helium burning under non-degenerate conditions, but its description is different from the degenerate case \citep[see e.g.,][]{Miller2022}, and consequently requires a different approach. Such a study is beyond the scope of this paper.}. It must be noted, though, that this relation also depends on chemical composition, for which Z can be taken as a proxy. Typically, this dependence is taken into account by computing individual relations for a given set of metallicity values, as done in  \citet{Rappaport1995} or \citet{Chen2013}.

What follows is to assume that the radius provided by the core mass -- radius relation is approximately equal to the Roche lobe radius of the sdB progenitor right before the end of mass transfer, the moment at which its mass is also close to the final sdB mass. In doing so, a relation that links the Roche lobe radius and the orbital separation \citep[e.g.,][]{Eggleton1983} can be used with Kepler's third law to derive the final form of the P-q relation. The reader might notice that at this point, an additional dependence on the sdB mass remains, from both Kepler's third law and the core mass -- radius relation. In general, if this value is unknown a priori, it is assumed that it takes the canonical value (though see \ref{sect:mpdist} for an analysis from a BPS perspective).

\subsubsection{Common Envelope}

Systems in which the sdB is born from a common envelope event present a comparatively more complex evolutionary history than their analogue from the stable mass transfer scenario. This is in part due to their evolutionary history, which encompasses a much wider variety of possible companions and types of mass transfer events pre-sdB formation, and in part due to the poorly understood common envelope process itself. Hence, we opt to focus on a particular case that has a semi-analytical step-by-step derivation provided by \citet{Zhang2021}: sdB + HeWD systems.

This case makes use of the following assumptions. First, the HeWD companion is born from a stable mass transfer event that can be modelled similarly to what was presented in section \ref{sect:stable}, though in this case the core mass would correspond to the HeWD mass. The expression for the orbital separation that can be obtained by equating the radius of the HeWD progenitor and the Roche lobe radius right before the end of mass transfer is not directly used in Kepler's third law but instead is considered as the initial separation at the beginning of the common envelope process. This implicitly assumes that no other processes have greatly impacted the orbital separation between the end of the first stable mass transfer event and the onset of the common envelope event. Then, it is assumed that the common envelope itself can be modelled in such a way that the initial separation can be directly related to the post-CE separation, such as in \citet{deKool1990}. These recipes usually add new variables like the envelope removal efficiency (labelled $\alpha_{\rm CE}$) and a structural parameter (referred to as $\lambda$). Either detailed modelling or simplified assumptions are then employed to estimate the values that these variables take, after which one can finally use the post-CE separation in Kepler's third law and find the sought P-q relation.

\section{Results}\label{sect:res}

The following sections contain our main results, presenting the numbers of the synthetic current-day sdB population in subsection \ref{sect:numbers}, an analysis of the P-q relation in subsection \ref{sect:pqres}, and the observed distributions of surface parameters (Kiel diagram), masses and orbital periods in both subsections \ref{sect:kiel} and \ref{sect:mpdist}.

\subsection{Current Day Population}\label{sect:numbers}

The details of the distribution of different companion types and formation channels at the current epoch are presented in Tables \ref{tab:companions} and \ref{tab:channels}, respectively. As can be initially interpreted from these, most of our results correspond to stars that belong to the thin disk and are born through stable mass transfer episodes (no common envelope). This can be understood if we consider the thin disk mass fractions defined by the Besan\c{c}on model used in this work (see Table \ref{tab:besancon}), and the age range defined by the same model for each Galactic component. In Figure 13 of Paper I we showed that most sdBs are born through stable mass transfer and that there is a preference for young (a few Gyr old) systems to become sdB progenitors. Thin Disk 2 and 3 clearly show this pattern, while Thin Disk 1 is still too young to produce sdBs in large numbers. It should not be a surprise that young systems are also correlated with more massive stars since these evolve faster and therefore form a helium core before their low-mass counterparts reach such an evolutionary stage. Ignoring extreme mass ratios, this also affects the average companion masses, as can be seen by the large and predominant number of sdBs with massive ($M \geq 1 M_\odot$) MS companions. This conflicts with observational samples that do not contain large fractions of sdB plus massive MS companions \citep[associated with what we have previously defined as early-type, e.g.,][]{Green2004, Vos2019, Schaffenroth2022} in large numbers. However, this does not invalidate our results, since the observational sample is known to be incomplete and biased {\em against} early MS companion types\footnote{Throughout this work, we make use of \textit{early-type} as a synonym of spectral type earlier than $\sim$F. The usage of \textit{late-type} also stems from this definition.}, owing to the contribution of these massive MS stars to the observed flux of the system and associated difficulties when using standard sdB classification techniques \citep[e.g.,][and references therein]{Dawson2024}. This motivates our choice of separating sdB systems based on their MS companion mass within Table \ref{tab:companions}. 

Systems in which the companion is in an evolutionary stage other than MS seem to be less likely for several reasons. The typical short $\sim 100$ Myr lifetime of the core helium burning phase corresponding to the sdB lowers the evolutionary overlap chance for all stellar types, while inherent low occurrence rates or short evolutionary stages of e.g., neutron stars, black holes, or Hertzsprung gap stars further contribute to the low-occurrence probability. This, on the other hand, does not represent a problem for white dwarf companions, since they correspond to the final evolutionary stage for most stars in the Galaxy \citep{Fontaine2001}. We can see that these systems represent a large fraction of the final sdB population, even greater than other non-MS stellar types combined.

\begin{table*}
\centering
\caption{Number of sdB systems in our synthetic population, separated  by Galactic component (rows) and companion type (columns). Stellar classes included in the \textit{Other Non-Remnant} column include sub-giants, giants, core helium-burning, and helium stars. Similarly, \textit{Other Remnants} is composed of oxygen-neon WDs, neutron stars and black holes. The top row for each Galactic component corresponds to the total number of sdBs found within it, while the bottom row shows our results limited to a 500pc radius around the Sun's location as defined in our model.}\label{tab:companions}
{\begin{tabular}{@{\extracolsep{\fill}}llllllll}
\toprule
Component & ${\rm MS}\leq 1M_\odot$&${\rm MS}> 1M_\odot$&Other Non-Remnant&HeWD&COWD&Other Remnants&Mergers\\
\hline
Thin Disk 1 & 0 & 0 & 0 & 0 & 0 & 40 & 0 \\
  & 0 & 0 & 0 & 0 & 0 & 0 & 0 \\
Thin Disk 2 & 32900 & 1861930 & 26620 & 10390 & 200390 & 42550 & 0 \\
  & 30 & 1770 & 10 & 20 & 300 & 20 & 0 \\
Thin Disk 3 & 17720 & 470240 & 37570 & 6510 & 133650 & 16330 & 0 \\
  & 20 & 510 & 20 & 0 & 110 & 10 & 0 \\
Thin Disk 4 & 35480 & 49330 & 20510 & 49170 & 8720 & 250 & 350 \\
  & 50 & 10 & 0 & 40 & 0 & 0 & 0 \\
Thin Disk 5 & 99270 & 43880 & 24300 & 112660 & 3450 & 0 & 1000 \\
  & 30 & 80 & 10 & 70 & 10 & 0 & 0 \\
Thin Disk 6 & 104310 & 25640 & 19150 & 74190 & 780 & 0 & 9480 \\
  & 90 & 20 & 10 & 50 & 0 & 0 & 0 \\
Thin Disk 7 & 182880 & 7480 & 18160 & 109280 & 3520 & 0 & 54100 \\
  & 110 & 10 & 10 & 50 & 0 & 0 & 40 \\
Bulge & 68420 & 5240 & 8720 & 36380 & 1940 & 0 & 20410 \\
  & 0 & 0 & 0 & 0 & 0 & 0 & 0 \\
Thick Disk & 150260 & 0 & 7440 & 98620 & 9930 & 0 & 66250 \\
  & 10 & 0 & 0 & 60 & 0 & 0 & 0 \\
Halo & 6000 & 0 & 0 & 4490 & 160 & 0 & 6330 \\
  & 0 & 0 & 0 & 10 & 0 & 0 & 0 \\
\hline
Total & 697240 & 2463740 & 162470 & 501690 & 362540 & 59170 & 157920 \\
  & 340 & 2400 & 60 & 300 & 420 & 30 & 40
\botrule
\end{tabular}}
\end{table*}

For completeness, we also show in Table \ref{tab:channels} the numbers associated with each formation channel discussed in Paper I. Even after applying the Besan\c{c}on model, the relevance of each channel remains almost unaltered from what we presented in our previous work. The most common formation channel is stable mass transfer, followed by sdBs being born as the direct result of a single common envelope event. More intricate cases as sdBs being born from mergers, after two common envelope episodes, or via the Early Common Envelope (E CE) are the most rare formation channels, though the latter would be rather similar to the stable mass transfer formation scenario.

Following the recent work by \citet{Dawson2024}, we were also motivated to study an analogue 500pc synthetic population of sdBs, shown as the bottom row in every component depicted in Tables \ref{tab:companions} and \ref{tab:channels}. For this purpose, the Sun's location has been defined as $R_0 = 8.2$ kpc and $z_0 = 0.025$ kpc \citep{BlandHawthorn2016}. In general, the relative relevance of companion types and different formation channels is a scaling of the results for the entire Galaxy that we have found so far, except for mergers. In our model, this happens since mergers take longer to occur than other formation channels, as shown by a comparison between the distributions shown in Figures 13 and 14 of Paper I. As such, mergers are associated with older populations in the model \citep[for another study with similar results see][]{Han2008}, and since by definition neither the Halo nor the Bulge contribute much to the local stellar density, a different trend for the merger scenario is visible after limiting the sample to a local 500pc around the Sun.

By constraining our population to be in line with the local sample, in principle, we find 3590 systems within our current-day local population, which is strikingly higher than the 178 hydrogen-rich sdBs found by \citet{Dawson2024}. This disagreement stems from several factors. Systems with early-type MS companions are incomplete in the observational sample, while they represent most of our synthetic sdBs. In the most extreme scenario, we can assume that they have not been observed and reduce our 500pc sample to 1190 sdBs, still $\sim 7 \times$ higher than the observed population. Another potential factor in the disagreement corresponds to our definition of what the \textit{current-day population} means. The 50 Myr uncertainty initially introduced is substantiated by the usual sdB lifetime, though it does not properly fit a comparison with the observational current-day sdB population. Arbitrarily lowering the uncertainty to $\pm$2.5 Myr results in reducing the total number of local sdBs to 2410, which is further lowered to 720 systems if we remove candidates with massive MS companions. This final result is still 4 times greater than the observed population and, though it could be improved by modifying our initial configuration, we consider that in doing so we would end up overfitting our model.

Suffice it to say, the mass range allowed for helium ignition stands as the parameter with the largest potential to directly change the computed numbers among the different BPS parameters explored in Paper I. On top of that, further customizing the star formation history within the Besan\c{c}on model would also affect our results, particularly when it comes to the number of blue (massive MS) stars if we consider a decaying exponential star formation rate within the Thin Disk components instead of our current constant implementation \citep{Aumer2009, Czekaj2014}. These different configurations are beyond the scope of this paper, though we present them as options in case better agreement is required. An additional point to make here is that our synthetic sdB population is potentially incomplete as a consequence of the cut on companion mass below 0.08 $M_\odot$, which is motivated by the limits on the evolutionary tracks available within \textsc{compas} but would exclude systems with low-mass companions \citep[e.g.,][]{Green2004, Schaffenroth2022}. While we cannot estimate the differences that this would cause numerically, we think that the numbers of sdBs with early-type MS companions would be reduced owing to the IMF, while at the same time increasing the numbers for systems with late-type companions.

\begin{table}
\centering
\caption{Same as Table \ref{tab:companions}, but this time columns correspond to the different evolutionary channels presented in Paper I.}\label{tab:channels}
{\begin{tabular}{@{\extracolsep{\fill}}llllll}
\toprule
Component&No CE&E CE&1 CE&2 CE &Mergers\\
\hline
Thin Disk 1 & 40 & 0 & 0 & 0 & 0\\
  & 0 & 0 & 0 & 0 & 0\\
Thin Disk 2 & 1942020 & 23800 & 196400 & 12560 & 0\\
  & 1840 & 10 & 270 & 30 & 0\\
Thin Disk 3 & 522530 & 7010 & 129410 & 23070 & 0\\
  & 540 & 0 & 120 & 10 & 0\\
Thin Disk 4 & 73090 & 630 & 89460 & 280 & 350\\
  & 30 & 0 & 70 & 0 & 0\\
Thin Disk 5 & 117380 & 1150 & 165030 & 0 & 1000\\
  & 110 & 0 & 90 & 0 & 0\\
Thin Disk 6 & 115450 & 780 & 107840 & 0 & 9480\\
  & 70 & 0 & 100 & 0 & 0\\
Thin Disk 7 & 162730 & 160 & 158430 & 0 & 54100\\
  & 80 & 0 & 100 & 0 & 40\\
Bulge & 68900 & 490 & 51310 & 0 & 20410\\
  & 0 & 0 & 0 & 0 & 0\\
Thick Disk & 145010 & 0 & 121240 & 0 & 66250\\
  & 0 & 0 & 70 & 0 & 0\\
Halo & 4070 & 0 & 6580 & 0 & 6330\\
  & 0 & 0 & 10 & 0 & 0\\
\hline
Total & 3151220 & 34020 & 1025700 & 35910 & 157920\\
  & 2670 & 10 & 830 & 40 & 40
\botrule
\end{tabular}}
\end{table}

Yet another relevant result from \citet{Dawson2024} corresponds to the birth rate. However, we once again stumble upon all the caveats previously mentioned, and additionally, we highlight that the final birth rate presented in \citet{Dawson2024} considers all hot subdwarf types, while here we only consider sdBs. Moreover, our model provides a complete census of all Galactic components, unlike the estimates of \citet{Dawson2024} that stem from the local population. As a consequence, our results are not directly comparable, but we provide some rough estimates related to the birth rate. First, if we consider all the 4,404,770 sdBs in our sample, a total Galactic volume\footnote{Here we have taken the volume of a cylinder with a radius equal to the maximum radius in our model (30 kpc), and a height equal to the total height (4 kpc) in our model.} of $\sim 1.1\times10^{13}\,\rm{pc}^3$, and a temporal baseline of 100 Myr (the defined time uncertainty), we arrive at a rate $\sim 4\times10^{-15}\,{\rm stars\,\,yr}^{-1}\,\rm{pc}^{-3}$. This number is remarkably similar to the $3.35^{+1.24}_{-0.77}\times10^{-15}\,{\rm stars\,\,yr}^{-1}\,\rm{pc}^{-3}$ from \citet{Dawson2024}, though our Galactic volume estimation is rather crude and the stellar density in our model is not uniform. Similar to what was explored for the total number of sdBs, an alternative is to estimate the local birthrate based on the synthetic 500\,pc sample limited to a time baseline of 5\,Myr. This time, the rate corresponds to $\sim 9.2\times10^{-13}\,{\rm stars\,\,yr}^{-1}\,\rm{pc}^{-3}$, while removing systems with massive MS companions results in $\sim 2.8\times10^{-13}\,{\rm stars\,\,yr}^{-1}\,\rm{pc}^{-3}$ instead. The local birth rate is much higher than the observational predictions, though this is most likely caused by the same problems discussed for the 500pc limited synthetic sdB population.

\begin{figure}
\centering
\includegraphics[width=1\linewidth]{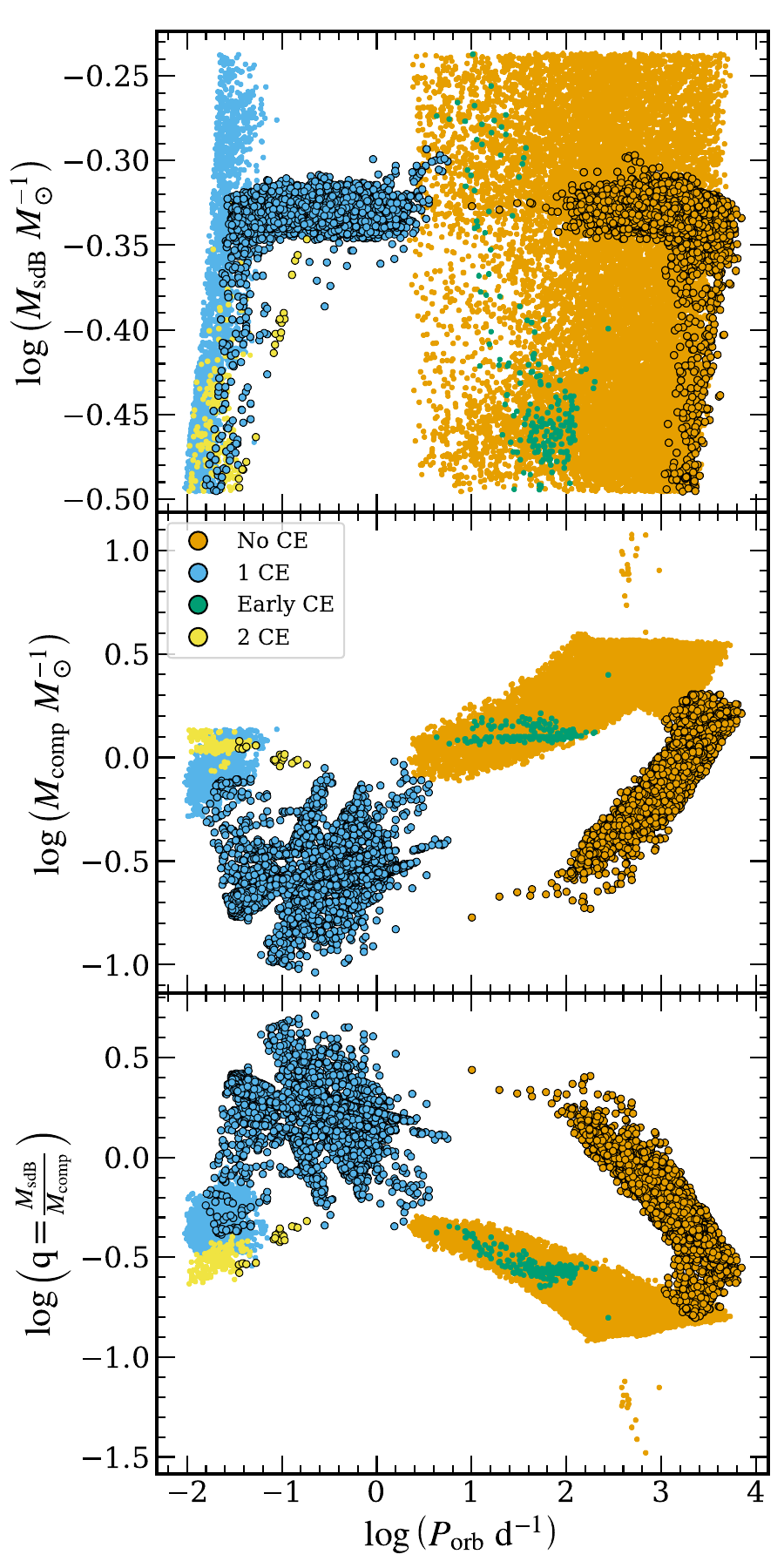}
\caption{The P-q relation as found in our BPS results. The logarithm of the orbital period (in days) is shown on the x-axis, while each panel from top to bottom shows the logarithm of the sdB mass, the companion mass, and the mass ratio in the y-axis. Markers with black edges represent sdBs born from progenitors that experience a flash during helium ignition.}
\label{fig:pq_all}
\end{figure}

\subsection{Insights on the P-q Relation}\label{sect:pqres}

\begin{figure}
\centering
\includegraphics[width=1\linewidth]{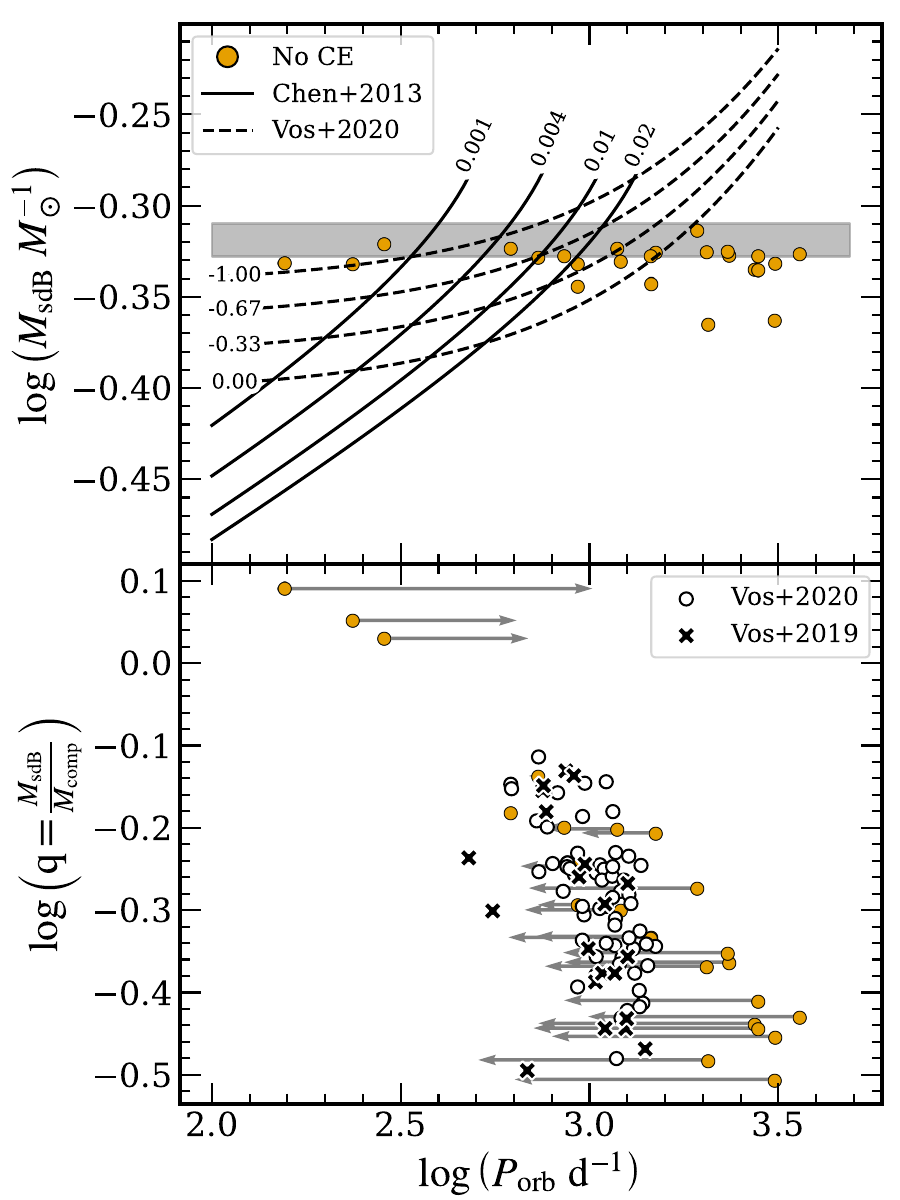}
\caption{Similar to Figure \ref{fig:pq_all}, but omitting the panel with companion masses and showing a narrower orbital period range. This figure only shows sdBs born from the stable mass transfer channel that have MS companions and computed radial velocities $K_{\rm MS} \geq 3\,\rm{km\,s}^{-1}$ and $K_{\rm sdB} \geq 5\,\rm{km\,s}^{-1}$, following similar limitations to the observational sample from \citet{Vos2019}. We have further limited the sample to candidates in a 500 pc radius. In the top panel, results of the period -- sdB mass relation for different Z values from \citet{Chen2013} are shown with the specific Z values shown at the right-most end of each black solid line, while the equivalent is done for several representative [Fe/H] values (at the left-most end of the dashed black lines) for the relation presented in \citet{Vos2020}. The grey area represents the \textit{canonical} sdB mass range 0.47 -- 0.49$M_\odot$ for reference. The bottom panel shows the P-q panel instead, with the observational results from \citet{Vos2019} as white circles, the population synthesis results from \citet{Vos2020} as crosses, and the population synthesis results from this work as orange circles. The arrows pointing away from the latter correspond to the orbital period computed using the theoretical approach described in section \ref{sect:stable}.}
\label{fig:pq_no_cee}
\end{figure}

\begin{figure}
\centering
\includegraphics[width=1\linewidth]{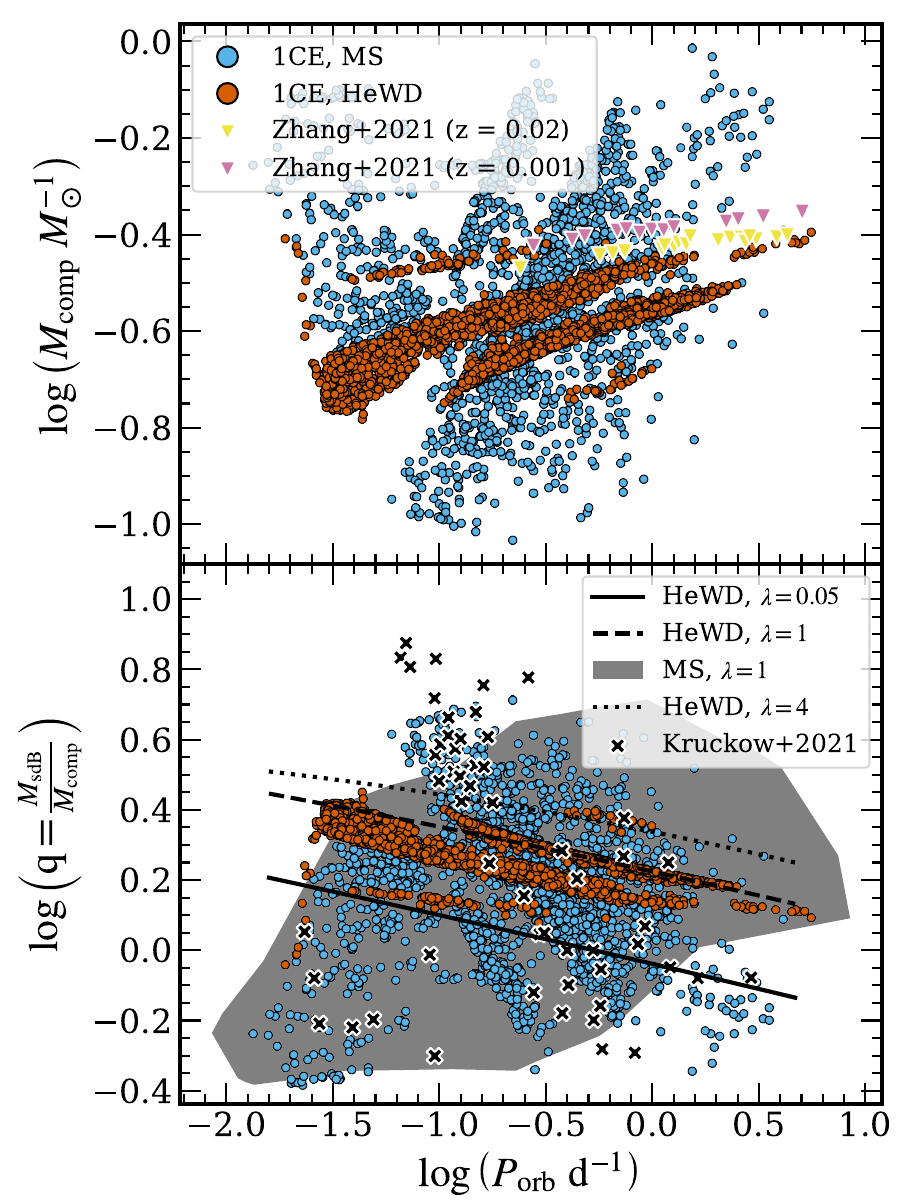}
\caption{Similar to Figure \ref{fig:pq_all}, but omitting the panel with sdB masses and focusing on sdBs born from systems that experience a single common envelope, the \textit{1CE channel} from Paper I. Blue markers indicate an MS companion, while orange is used to show HeWD companions. The top panel shows the orbital period -- companion mass relation, where we also show the results from \citet{Zhang2021} using detailed models, in general agreement with our population synthesis results. The bottom panel shows the population synthesis results in the P-q space, with the inclusion of observational estimates for sdBs in potential post-common envelope systems compiled by \citet{Kruckow2021}. The different lines correspond to different assumptions for the value of $\lambda$ in systems with HeWD companions, following section \ref{sect:pq}, while the shaded area corresponds to the area covered by applying a similar process to MS companions with $\lambda = 1$.}
\label{fig:pq_1_cee}
\end{figure}

We show the P-q plane as seen from our rapid BPS approach in Figure \ref{fig:pq_all}. From top to bottom, different forms of the relation are presented: P-$M_{\rm sdB}$, P-$M_{\rm comp}$ and P-q. It might seem like there is no clear and useful relation between the orbital period and the stellar masses in the system since populated regions do not seem to follow any easily describable trend. This is somewhat improved by making a distinction between systems where the sdB has been born from a progenitor that ignited helium in a flash (circles with black edges in the figure) and those which do not. Indeed, a subset of long-period systems with such characteristics seems to follow a linear trend in both the P-$M_{\rm comp}$ and P-q planes, albeit with non-negligible dispersion. This corresponds to one of the most studied subsets \citep[i.e.,][]{Chen2013, Vos2020} and is presented in further detail in Figure \ref{fig:pq_no_cee}. A similarly studied relation for short-period sdB systems with HeWD companions \citep[e.g.,][]{Zhang2021} is depicted in Figure \ref{fig:pq_1_cee}, though much more intricate than what is seen for long-period systems. Details for these subsets of the P-q relation are further discussed below.

Binary systems composed of an sdB and a low-mass MS companion in long-period orbits \citep[$\geq 400$ d,][]{Chen2013,Vos2019, Vos2020} have been known to follow a simple polynomial relation between the sdB mass and the orbital period, with an additional dependence on the chemical composition of the system that helps to explain the observed spread \citep{Vos2020}. Since these systems are born from progenitors that ignite helium in a flash, it is expected that their masses will be tightly concentrated around canonical mass (as seen in Figure \ref{fig:pq_no_cee} and discussed in section \ref{sect:mpdist}), though some variations might be observed for systems that are close to the mass limit for smooth helium ignition (e.g., the spread in dark-edged orange circles below $\log{M_{\rm sdB}\sim-0.35}$ in the top panel of Figure \ref{fig:pq_all}). This relation is also recovered from the results of our population synthesis, and a graphic comparison against the theoretical approach described in section \ref{sect:stable} as well as other studies \citep{Chen2013,Vos2019,Vos2020} is shown in Figure \ref{fig:pq_no_cee}. We must clarify that the number of stars in this figure is rather low, as we have imposed observationally-motivated constraints: radial velocity limits from \citet{Vos2019} and systems located within 500pc of the Sun. Additionally, in the bottom panel of Figure \ref{fig:pq_no_cee}, arrows point towards the expected theoretical orbital period computed following the ideas described in section \ref{sect:stable}, for each one of the mass ratios of the systems generated through our BPS study. These re-computed periods seem to be in better agreement with observational data and results from \citet{Vos2020}, though the source of discrepancies can be also linked to the particulars of the evolutionary schemes available within \textsc{compas}. That is, it is also possible to improve agreement by tuning the BPS configuration, but the current results cover a remarkably similar parameter space and therefore we do not consider it warranted to impose any further modifications. 

The case of sdBs in short periods with HeWD companions is more complicated to assess and make use of, in contrast to the sdB + MS stars formed after a stable mass transfer scenario. While its potential usefulness in constraining evolutionary processes such as the common envelope cannot be denied \citep{Zhang2021}, population synthesis results from this work seem to hint at a considerable spread in the P-q relation, owing to the stellar structure-related parameter $\lambda$ (see Figure \ref{fig:pq_1_cee}, bottom panel). While metallicity also plays a role in the spread of this relation, the biggest factor seems to be the evolutionary status of the progenitor and subsequent $\lambda$ value. Further, we have made the assumption of $\alpha = 0.2$ based on observational results \citep{Zorotovic2010}, but its actual value and its treatment in numerical simulations are still a topic of debate and add to the global uncertainty \citep{Iaconi2019}, since the usual prescriptions used to describe the common envelope process heavily rely on both $\alpha$ and $\lambda$ \citep{Roepke2023}. Finally, from an observational standpoint, the evident overlap with sdBs paired to MS companions also increases the difficulty of establishing a clear P-q relation such as the one that was discussed for long orbital period systems. Since $\lambda$ is related to the sdB progenitor's structure, the nature of the companion does not greatly affect the parameter space populated by the system after the common envelope. In the bottom panel of Figure \ref{fig:pq_1_cee}, we show the computed outcomes of common envelope episodes between sdB progenitors and their MS companions by using the \citet{deKool1990} prescription with $\alpha = 0.2$ and $\lambda = 1$. The result (grey area) is, for the most part, indistinguishable from the sdB+HeWD systems. While in principle it is possible to identify the latter through ellipsoidal deformation, or similarly identify cool companions by the reflection effect \citep[e.g.,][]{Schaffenroth2022}, deciding whether the WD corresponds to a helium or CO one and properly constraining the companion masses represent additional challenges. This is further highlighted in Figure \ref{fig:pq_1_cee}, where we have plotted the observational post-common envelope sdB sample from \citet{Kruckow2021}, which shows no clear preference for any system configuration.

\subsection{The Kiel Diagram}\label{sect:kiel}

\begin{figure*}
\centering
\includegraphics[width=1\linewidth]{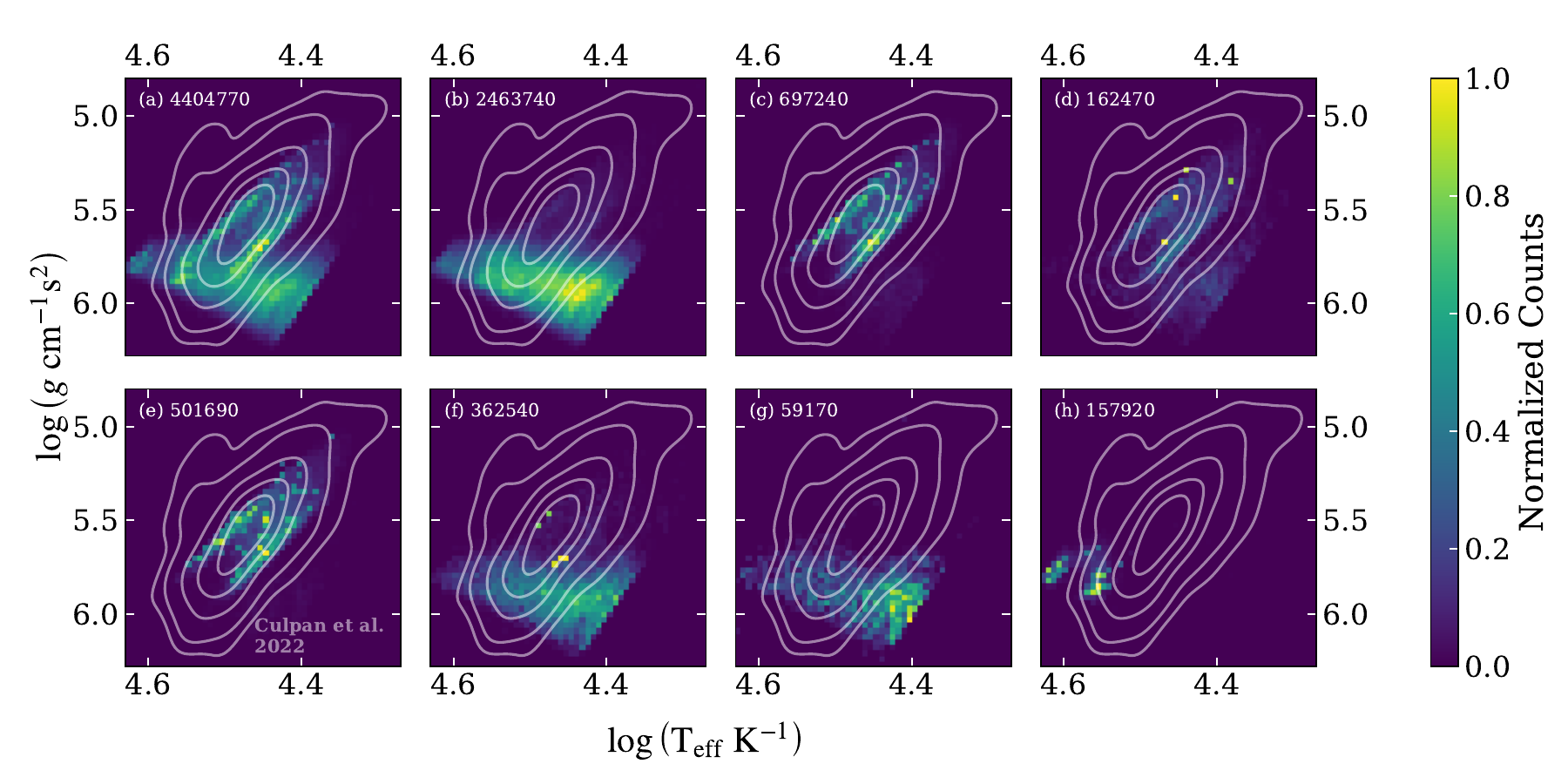}
\caption{All panels depict the Kiel diagram for the synthetic binary population results and observations. The former are represented through 2D histograms, where the more yellow (dark blue) a bin is, the more (less) systems it contains. The white contours correspond to a Kernel Density Estimation applied to the observational sample from \citet{Culpan2022}, where the outermost contour corresponds to 90\% iso-proportion of the density (i.e., 90\% of the probability mass lies inside this contour); the next-most inner contour corresponds to 80\%, and when going inwards further the iso-proportion represented by each contour decreases in steps of 20\%. Panels are defined as follows: \textbf{(a)} all systems, \textbf{(b)} systems with MS companions where $M_{\rm comp} > 1 M_\odot$, \textbf{(c)} systems with MS companions where $M_{\rm comp} \leq 1 M_\odot$, \textbf{(d)} systems with non-remnant companions (sub-giants, giants, core helium-burning, and helium stars; or equivalently stellar types 2 to 9 in the \citeauthor{Hurley2000} \citeyear{Hurley2000} scheme), \textbf{(e)} systems with HeWD companions, \textbf{(f)} systems with COWD companions, \textbf{(g)} systems with other remnants as companions \citep[oxygen-neon WDs, neutron stars and black holes; stellar types 11 to 14 in][]{Hurley2000}, and \textbf{(h)} double HeWD mergers. The number in the top-left corner of each panel corresponds to the total number of systems in the 2D histogram.}
\label{fig:kiel_cont}
\end{figure*}

Figure \ref{fig:kiel_cont} shows the comparison between our population synthesis results and the observational data of \citet{Culpan2022}. A first evident feature of our results is how they overlap with a considerable fraction of the observational sample, even though the \citet{Bauer2021} models contain hydrogen-rich envelope masses only up to $3\times10^{-3}M_\odot$, one order of magnitude lower than the upper limit of $\sim2\times10^{-2}M_\odot$ found in the literature \citep[e.g.,][]{Heber2024}. Additionally, the figure shows an interesting agreement between the clustering area of the observational sample and that of sdB systems with late-type MS companions (panel \textit{c}), in line with the observational constraints presented by e.g., \citet{Dawson2024}. A similar trend can be seen for HeWD companions in panel \textit{e}, due to the sdBs in these systems also being associated with low-mass progenitors that ignited helium in a flash, resulting in masses tightly concentrated around the canonical value. These results  support the idea of using a mass close to canonical when studying sdBs in such configurations without a properly constrained mass value. Although the overlap is not perfect, the differences do not necessarily represent a problem with the population synthesis implementation and are rather caused by the chosen envelope mass distribution, as discussed in section \ref{sect:synth}. 

It must be noted that the remaining panels highlight the potential problems of using canonical mass in other configurations, as the preferred locus in the Kiel diagram for panels \textit{b}, \textit{d}, \textit{f} and \textit{g} is different from what can be seen in panels \textit{c} and \textit{e}, as well as the observational trend. In particular, temperature spread in an sdB sample at a fixed surface gravity value is mainly associated with both age and mass, though the $\sim0.2$ dex temperature coverage in panels \textit{b}, \textit{f} and \textit{g} can only be explained by sdB masses covering the range $\sim 0.3$--$0.5~M_\odot$ within the population. Moreover, if the assumption discussed in section \ref{sect:numbers} about incompleteness within the observed population of sdB plus early-type MS companions is incorrect, then that would mean that the observed distribution should be closer to the 2D histogram shown in panel \textit{b}, which is not the case. A final possibility is that our adopted hydrogen-rich envelope description is incomplete, which is something we plan on analysing in future work.

A distinct scenario is shown in panel \textit{h}, where the sdB candidates formed after the merger of two HeWDs cluster at higher temperatures than the clustering area of the observed population at similar surface gravity values. Though an interesting result in itself, it is also noticeable that the observational sample presents a similar concentration that deviates from the main distribution, although at lower surface gravity. While potentially related, in Paper I we mentioned that our merger candidates need further analysis since their evolution is estimated by using the total mass of the system as an input for the hot subdwarf evolution prescription adapted from \citet{Bauer2021} instead of using detailed merger models, and therefore we cannot draw any specific conclusions. Particularly, the evolutionary tracks for hot subdwarfs born from mergers \citep[e.g.,][]{Zhang2012} differ from our implementation of the \citet{Bauer2021} models and could result in a different clustering area within the Kiel diagram.

\begin{figure}
\centering
\includegraphics[width=1\linewidth]{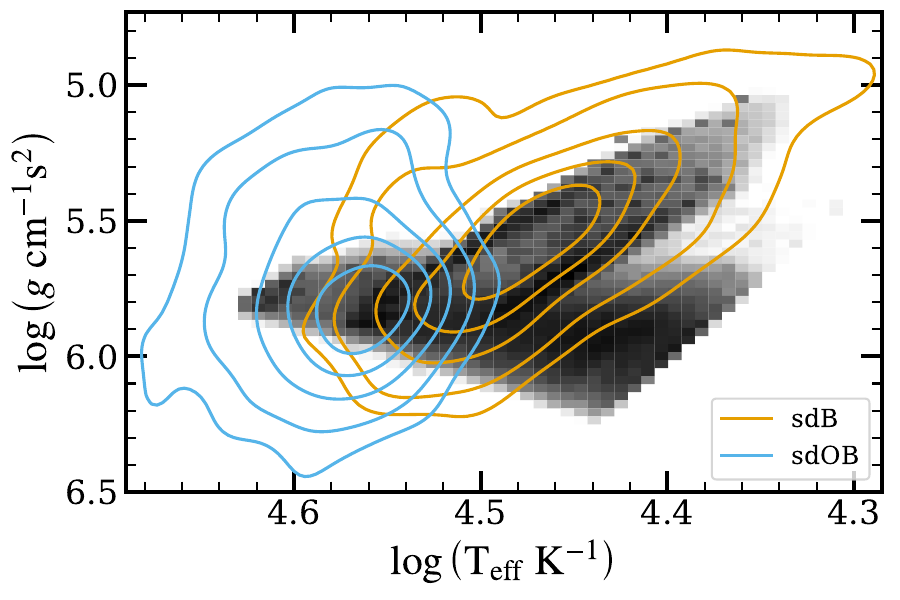}
\caption{The same content as panel \textbf{(a)} from Figure \ref{fig:kiel_cont} is shown, but using a different shade to highlight the distributions corresponding to the observational sample of sdBs and sdOBs from \citet{Culpan2022}. Note that there is an overlap at the hot end of the BPS results (greyscale 2D histogram) with both hot subdwarf classes (O and B) shown. The opposite is true for the high-gravity, cool end of the BPS sample: there is no evident overlap between the simulated population and the observational distribution.}
\label{fig:kiel_zoom}
\end{figure}

The discussion around non-canonical hot subdwarfs in panels other than \textit{a}, \textit{c}, and \textit{e} naturally takes us to consider whether all these stars are sdBs or a fraction of them corresponds to other hot subdwarf classes. In Figure \ref{fig:kiel_zoom} we present a visualization that includes the typical clustering areas of all sdBs from our population synthesis, and the observational results from \citet{Culpan2022} for both sdB and sdOB\footnote{Note that type O hot subdwarfs have not been considered due to our population being constituted by core-helium burning subdwarfs only.} \citep[hot subdwarfs of OB spectral type, e.g.,][]{Moehler1990,Geier2017a}. This can be understood as a zoom-in of panel \textit{a} from Figure \ref{fig:kiel_cont}, with added contours for sdOBs. An evident result is the overlap at $\log{g}\sim5.8$ and $\log T_{\rm eff}\sim 4.56$, which hints at the possibility that our BPS results include not only sdBs but also sdOBs. A limitation of our results, however, is that they do not include information related to surface composition, meaning that we cannot make a direct assessment of the possible sdOB "contamination" in our sample. Still, it is possible to evaluate the inclusion of sdOB stars from \citet{Dawson2024} into the target number of hot subdwarfs within 500pc. This results in 243 sdB+sdOB stars in the observational sample, approximately one-third of our best estimate (720 systems) from section \ref{sect:numbers}. 

Further "optimistic" corrections can be made by ignoring results that do not overlap with observational results, i.e., removing the cool and compact end of the BPS distribution. Completely removing these results in 420 systems instead, about 1.7 times the observational sdB+sdOB numbers within 500 pc. Since we cannot undeniably link this correction to known constraints (e.g., as done for early MS companions), we refrain from labelling it as a reliable calculation. However, a potential reason that could support these corrections lies within the role of the $\lambda_{\rm CE}$ parameter, estimated through the prescription by \citet{Xu2010,Xu2010a} implemented in \textsc{compas}. This prescription does not include the contribution of recombination to the binding energy of the envelope, which potentially results in more systems surviving (not merging) when computing the outcome of a common envelope episode \citep[e.g.,][]{Marchant2021}. A considerable fraction of the non-canonical sdBs from the BPS results with companions other than MS stars (panels \textit{f} and \textit{g}) are born through common envelope episodes, and would potentially be removed from the sample under a different $\lambda_{\rm CE}$ prescription that includes the contribution of recombination to binding energy. However, at the same time, such a choice would probably increase the number of mergers, implying that a thorough analysis is required.

Finally, we note that the BPS overestimation of systems in short periods (product of common envelope episodes) is not exclusive to hot subdwarfs, but has also been reported in studies of AM CVn systems \citep[e.g.,][]{Roelofs2007,Rodriguez2025}.

\subsection{The Mass and Period Distributions}\label{sect:mpdist}

\begin{figure}
\centering
\includegraphics[width=1\linewidth]{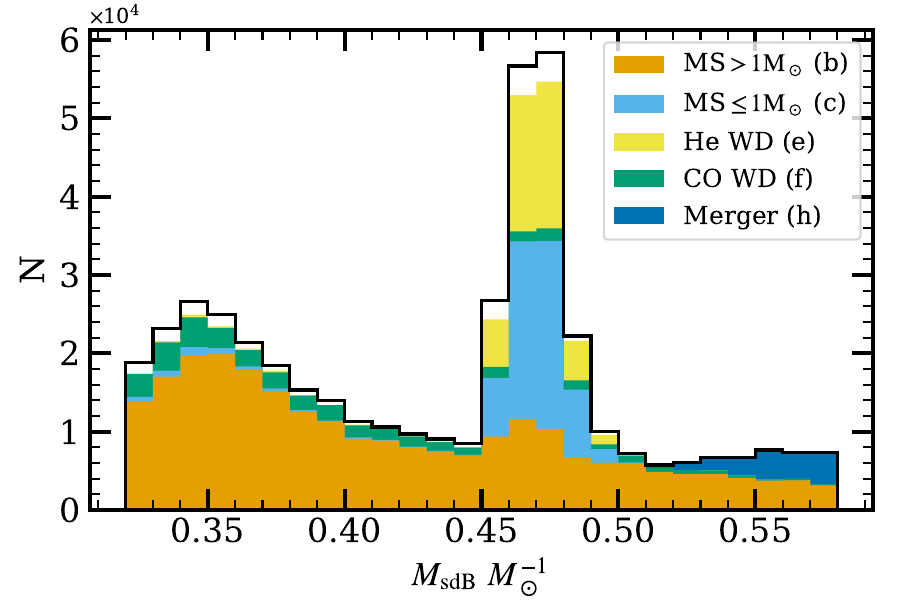}
\caption{Mass distribution of the current-day sdB synthetic population. The outer black solid line represents the total amount of systems per bin, while the different colours represent individual contributions by different companion types (except for mergers, that do not have companions). Corresponding designations from Figure \ref{fig:kiel_cont} are included in parentheses within the labels. The leftover white space in each bin corresponds to all other companion types, as shown in Table \ref{tab:companions}.}
\label{fig:sdb_mass}
\end{figure}

We present the mass distribution of our Galaxy-like population in Figure \ref{fig:sdb_mass}. This corresponds to the entire sample, with different colours representing companions of different stellar classes. For the most part, the canonical mass peak is composed of sdBs with low-mass MS or HeWD as companions. Another peak at $\sim 0.35 M_\odot$ can be seen, though in this case it is created by early-type MS companions and it has not been confirmed observationally, as discussed in section \ref{sect:numbers}. Still, the idea of using a canonical mass when studying sdBs seems to only be justified in the particular cases of sdB plus HeWD or low-mass MS systems, and even in this case there is about $\pm0.02\,M_\odot$ uncertainty. Further, sdBs paired with COWD stars seem to be almost evenly distributed for most of the mass range being presented, while mergers show a clear preference for larger sdB masses due to their different formation mechanism. These characteristics should be taken as a sign of caution against assuming a fixed mass value for all observed hot subdwarfs, in particular sdBs.

While we present the entire mass range of results from our synthetic populations, we note that in Paper I we imposed mass limits owing to the availability of hydrogen-rich envelope models for sdBs with $M \geq 0.58\,M_\odot$, and therefore the distribution is incomplete in this regard. However, the number of sdBs per mass bin should only decrease with increasing mass (after canonical mass), considering the IMF preference for low-mass stars \citep[e.g.,][]{Hopkins2018}. This might be affected by the distribution of sdBs produced by mergers, though in that case, it would not be relevant for studies focusing on binary systems where the canonical mass assumption is often used to estimate the companion's nature and minimum mass. Additionally, as commented previously, our sample is missing systems where the companion to the sdB is below the hydrogen burning limit since we avoided extrapolating the models available in \textsc{compas}. If these systems follow the trends that we have found so far, they would further contribute to the canonical mass peak without changing the shape of the synthetic mass distribution otherwise.

\begin{figure}
\centering
\includegraphics[width=1\linewidth]{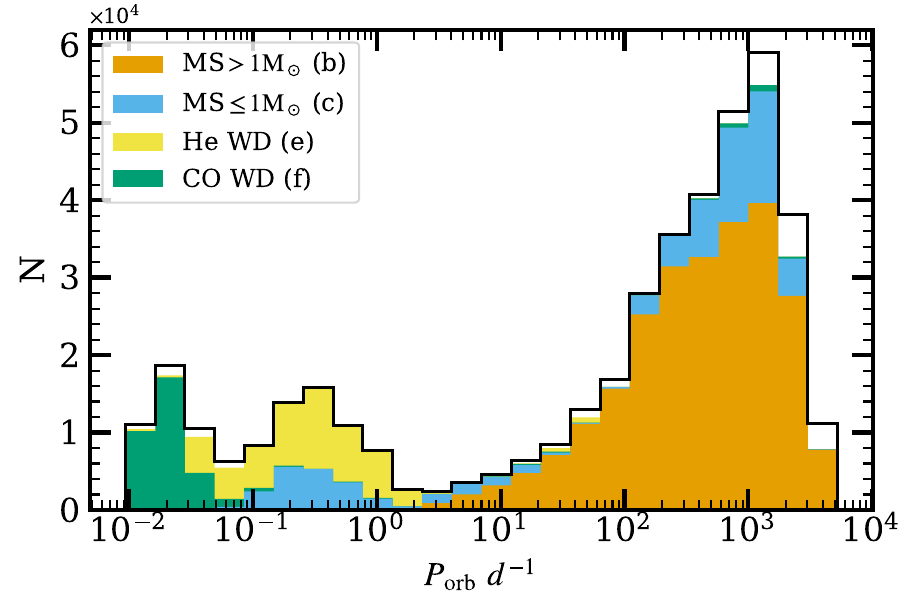}
\caption{The depicted orbital period distribution follows the concepts described in Figure \ref{fig:sdb_mass}. Mergers are not included for obvious reasons.}
\label{fig:sdb_p}
\end{figure}

The period distribution shown in Figure \ref{fig:sdb_p} also presents a result that differs from observations since it implies that the majority of sdBs are in long-period orbits. Once again we can argue that whether this result is real or not heavily depends on the existence of sdB plus early-type MS companions in large numbers. The discussion then follows what was already addressed in section \ref{sect:numbers}. 

Although expected from the detailed description of the different sdB formation channels \citep{Han2002,Rodriguez2024}, Figure \ref{fig:sdb_p} also provides the companion type probability based on the orbital period. For example, the shorter the orbital period, the more likely that the companion is a WD, with COWDs being found at the extreme of the distribution. While useful, the picture might be incomplete, as we are still missing non-WD companions with masses below the hydrogen burning limit, which would alter the distribution below $\sim1$ day periods \citep[e.g.,][]{Schaffenroth2022}.

%%%

\section{Conclusions}\label{sect:conc}

In this work, we have analyzed the Galactic population of sdB stars from a BPS perspective. We noticed that either a normal or log-normal envelope mass distribution based on the fits to the models of \citet{Bauer2021} better reproduce the observational results from \citet{Culpan2022} for sdBs, based on the assumption that most of the observational results from that sample correspond to sdBs with masses around the canonical value. Regarding this last concept, the \textit{canonical mass}, we found that only specific configurations (sdB + HeWD, sdB + late-type MS star) are highly likely to contain sdBs with such masses, while caution is advised when assuming a fixed canonical value for the mass of sdBs in other configurations. Further, our synthetic sample predicts a large number of sdB plus early-type MS companions ($M>1\,M_\odot$) which is not consistent with the available data from the literature, however, the known population suffers from incompleteness \citep[e.g.,][]{Dawson2024}. Removing these systems provides a better agreement between our synthetic population and the observational sample, particularly in terms of the distribution within the Kiel diagram (Figure \ref{fig:kiel_cont}).

We also find a P-q relation for long-period systems that is, in general, consistent with what has been found from observations \citep[e.g.,][]{Vos2020}. This indicates that rapid population synthesis is still a useful tool for sdB studies. In a similar vein, we have attempted to shed some light on what would be expected for an equivalent relation between the orbital period of sdB plus HeWD systems and their mass ratios. Our results indicate, however, that overlap with sdB plus MS companions as well as uncertainties associated with the common envelope process make this relation much harder to analyse from observational data than what has been done for long-period systems.

While we have attempted to provide the best possible estimate for the sdB population through a BPS study, we are also aware of a set of limitations. First, the number of sdBs found within 500pc is too large when compared to the sdBs found by \citet{Dawson2024}, even after removing the systems with early-type MS companions. Potential reasons for this discrepancy are the helium ignition range for systems below the expected mass at helium flash (we have kept a 5\% threshold from Paper I), the "contamination" caused by sdOBs in our synthetic sample, the effects of recombination in the computation of $\lambda_{\rm CE}$, and our implementation of the Besan\c{c}on model (assuming a constant SFR within each Galactic component, and using the \citeauthor{Czekaj2014} \citeyear{Czekaj2014} local densities to estimate the corresponding mass fractions). Also, our choice for envelope mass distribution has improved the agreement with the observed sample, but it lacks a solid physical argument. Additional questions related to these envelopes include their metallicity dependence, behaviour at larger sdB masses and modelling in the case of sdBs born from mergers. We intend to explore some of these in future work.

\section*{Acknowledgements}

We thank the anonymous referee for their valuable comments. NRS thanks Harry Dawson and Alexey Bobrick for helpful discussion and insight. AJR acknowledges financial support from the Australian Research Council under award number FT170100243.

\section*{Software}
The figures and methods shown throughout this work made use of the following set of software: \textsc{python} 3.10, \textsc{compas} \citep[Team COMPAS:][]{Riley2022}, \textsc{matplotlib} \citep{Hunter2007}, \textsc{scipy} \citep{Virtanen2020}, \textsc{numpy} \citep{Harris2020}, \textsc{astropy} \citep{astropy2013,astropy2018,astropy2022}, \textsc{pandas} \citep{pandas2010,pandas2022}, \textsc{h5py} \citep{h5py2014}, \textsc{seaborn} \citep{Waskom2021}, and \textsc{topcat} \citep{topcat2005}.

%%%%%%%%%%%%%%%%%%%%%%%%%%%%%%%%%%%%%%%%%%%%%%%%%%
\section*{Data Availability}
 
All data will be available upon reasonable request to the corresponding author.



\bibliographystyle{mnras}
\bibliography{main} % if your bibtex file is called example.bib

% Don't change these lines
\bsp	% typesetting comment
\label{lastpage}
\end{document}